\documentclass[final,5p,times]{elsarticle}
\usepackage{graphicx} 
\usepackage{dcolumn}  
\usepackage{bm}       
\usepackage{hyperref} 
\usepackage{physics}
\usepackage{amsmath}
\usepackage{amssymb}  
\usepackage[utf8]{inputenc} 
\usepackage{xcolor}

\journal{Materials Today Physics}

\begin{document}

\begin{frontmatter}

\title{Thermally induced band gap increase and high thermoelectric figure of merit of n-type PbTe}

\author[label1,label2]{Jiang Cao\corref{cor1}}
\cortext[cor1]{Corresponding author}
\ead{jiang.cao@njust.edu.cn}
\author[label2]{Jos\'e D. Querales-Flores}
\author[label2,label3]{Stephen Fahy}
\author[label2]{Ivana Savi\'c \corref{cor2}}
\cortext[cor2]{Corresponding author}
\ead{ivana.savic@tyndall.ie}
\address[label1]{School of Electronic and Optical Engineering, Nanjing University of Science and Technology, Nanjing 210094, China}
\address[label2]{Tyndall National Institute, Dyke Parade, Cork T12 R5CP, Ireland}
\address[label3]{Department of Physics, University College Cork, College Road, Cork T12 K8AF, Ireland}

\begin{abstract}

Unlike in many other semiconductors, the band gap of PbTe increases considerably with temperature. We compute the thermoelectric transport properties of n-type PbTe from first principles including the temperature variation of the electronic band structure. The calculated temperature dependence of the thermoelectric quantities of PbTe is in good agreement with previous experiments when the temperature changes of the band structure are accounted for. We also calculate the optimum band gap values which would maximize the thermoelectric figure of merit of n-type PbTe at various temperatures. We show that the actual gap values in PbTe closely follow the optimum ones between 300~K and 900~K, resulting in the high figure of merit. Our results indicate that an appreciable increase of the band gap with temperature in direct narrow-gap semiconductors is very beneficial for achieving high thermoelectric performance.

\end{abstract}


\end{frontmatter}

\section{Introduction}
\label{sec:intro}

Thermoelectric (TE) materials convert reversibly thermal energy to electrical
energy. The TE figure of merit is defined as $zT=\sigma S^2 T / \kappa$, where  $\sigma$ is the 
electrical conductivity, $S$ is the Seebeck coefficient, $T$ is the temperature, and $\kappa$ is the thermal 
conductivity~\cite{Snyder2008}. 
The efficiency of TE refrigerators and power generators increases with larger values of $zT$.
At higher temperatures, minority carriers caused by thermal activation contribute negatively 
to Seebeck coefficient and decrease $zT$~\cite{Rosi1959}. An increasing band gap with temperature can suppress these 
bipolar effects and increase $zT$. Such band gaps would also lead to larger effective masses and Seebeck coefficient at higher $T$ in direct gap semiconductors~\cite{Kane1957}.

Previous model calculations have determined the ideal values of band gaps that would result in maximal $zT$ values  for semiconductors with indirect and direct gaps and relatively parabolic bands~\cite{CHASMAR1959,Sofo1994,Mahan1989},
For direct gap semiconductors, if the band gap $E_G$ is smaller than $6 k_B T$, where $k_B$ is the Boltzmann constant, it was found that $zT$ decreases when $E_G$ decreases due to the presence of minority carriers. 
If $E_G>10 k_B T$, $zT$ may decrease or increase with increasing $E_G$ depending on the dominant electron scattering 
mechanism in the material. However, for real TE materials, where different 
scattering mechanisms compete, no previous work has calculated the optimum band gap values
at different temperatures from first principles.

 In most semiconductors, the band gap decreases with increasing temperature. A particularly interesting exception
 is the group of lead chalcogenides (PbS, PbSe, PbTe), whose direct narrow gap increases with increasing
 temperature~\cite{Cardona2005, Snyder2013, Gibson1952}. A positive temperature coefficient of the band gap, $\partial E_G/\partial T$,
 could
 lead to the optimum gap values in a wider temperature range than if this coefficient is negative. The band gap of PbTe
 increases from 0.19~eV at 30~K to 0.38~eV at 500~K~\cite{Snyder2013,Gibson1952} ($\partial E_G/\partial T
 \approx 4.7 k_B$). Recent calculations suggested that this peculiar increase of $E_G$ with $T$ may be correlated with large anharmonicity~\cite{Yu2018}. Our recent first-principles work revealed that this effect stems from the Debye-Waller and thermal expansion contributions to the temperature renormalization of PbTe's gap~\cite{Querales2019}. 
 
First-principles calculations of thermoelectric transport quantities are typically
carried out using the electronic band structure obtained from density functional or higher level theories,
where the band shifts due to temperature are not accounted for \cite{Scheidemantel2003,Singh2010,Sohier2014,Gunst2016,Bonini2016,Song2017,Li2018,Ponce2018,Bernardi2018}.
For example, recent {\it ab-initio} calculations of TE transport in PbTe used a fixed band gap value at all temperatures~\cite{Song2017}. In contrast, an earlier study indicated the importance of accounting for the temperature variations of the band structure when modelling TE transport in PbTe~\cite{Bilc2006,Mahanti2010}. 
Our previous work on this subject~\cite{Cao2018} focused on the electronic mobility of n-type PbTe up to room temperature, where this effect is not so prominent.

In this work, we calculate all thermoelectric transport properties of PbTe from first principles,
explicitly accounting for the temperature renormalization of the electronic band structure.
We find much better agreement between our results and experiments when we include the temperature induced changes of the band structure. The ideal band gap values that would maximize $zT$ at different temperatures have also been obtained.
The optimum gap values vary from $10 k_B T$ at 300~K to $6.5 k_B T$ at 900~K, due to the competition between longitudinal optical and acoustic phonon scattering at high doping concentrations and temperatures.  
The actual band gap values of PbTe are very similar to the optimum ones in the whole
temperature range. Therefore, the sizeable positive temperature coefficient of the narrow band gap in PbTe  
enables its high $zT$ values over a large $T$ range.
This work suggests that efficient thermoelectric materials with a broad working temperature range
may be found among direct narrow gap semiconductors with relatively large positive values of $\partial E_G/\partial T$ (several $k_B$).

\section{Method}
\label{sec:method}

\subsection{Temperature-dependent band structure}

PbTe is a direct narrow gap semiconductor with the gap located at four equivalent L points.
The L valleys give the largest contribution to the electronic conduction and TE transport in n-type PbTe~\cite{Cao2018}. At higher temperatures, the energies of the valence band maxima at $\Sigma$ become similar to those of the valence band maxima at L. We neglect this effect in the present calculations. 
The electronic band structure near the L point in PbTe is well described using the two-band Kane model derived from  $\bm{k}\cdot\bm{p}$ theory~\cite{Kane1957}.
The energy dispersion  in the two-band Kane model is non-parabolic, and for the band extrema at 
the L point satisfies the relation:
\begin{equation}
  \label{eq:kanemodel}
  \frac{\hbar^2}{2} \left( \frac{k^2_\parallel}{m^*_\parallel} +
    \frac{k^2_{\perp}}{m^*_{\perp}} \right) = E \left(1+\frac{E}{E_G} \right) ,
\end{equation}
where $E_G$ is the direct band gap, $k_\parallel$ and $k_\perp$ are the 
components of the wavevector parallel and perpendicular to the $\rm \Gamma$-L 
direction, respectively, and ${m_\parallel^*}$ and $m_\perp^*$ are the parallel and 
perpendicular effective masses, respectively. In the two-band Kane model, the valence 
band (VB) is a mirror image of the conduction band (CB) with the dispersion given as $(-E_G-E)$.

The band gap and the effective masses for the ground state of PbTe are obtained from 
density functional theory (DFT) calculations, using the Vienna \textit{ab-initio} simulation package 
(VASP)~\cite{KRESSE1996}. We use the screened Heyd-Scuseria-Ernzerhof (HSE03) hybrid  functional~\cite{Heyd2003,Heyd2004} and include the spin-orbit coupling (SOC), which correctly reproduces the experimental values of the band gap and effective masses at low temperatures~\cite{Murphy2018,Cao2018}. The basis set for the one-electron wave functions is constructed with the Projector Augmented Wave (PAW) method~\cite{Kresse1999}. For the PAW pseudopotentials, we include 
the $5d^{10}6s^2 6p^2$ states of Pb and $5s^2 5p^4$ states of Te as the valence states. A cutoff energy of $18.4$~Ha and a 8$\times$8$\times$8 $\bm{k}$-mesh are used for the electronic band structure of PbTe. 

The temperature induced renormalization of the electronic band gap is due to thermal expansion and electron-phonon coupling~\cite{Tsang1971,Cohen1975}. We calculate the gap variation due to thermal expansion using the lattice constant values that account for thermal expansion, and computing the corresponding band gap change using DFT~\cite{Querales2019}. We compute the temperature dependence of the lattice constant using lattice dynamics from first principles, as explained in Ref.~\cite{Querales2019}. The electron-phonon renormalization of the band gap is calculated using the Allen-Heine-Cardona theory~\cite{Allen1976,Allen1981,Allen1983}
and its density functional perturbation theory (DFPT) implementation in the {\sc ABINIT} 
code~\cite{abinit1,Gonze2016}. We use the local density approximation 
(LDA)~\cite{Caperley1980,Perdew1981} and Hartwigsen-Goedecker-Hutter norm-conserving pseudopotentials~\cite{Hartwigsen1998}
with the $6s^2 6p^2$ states of Pb and $5s^2 5p^4$ states of Te explicitly included in the 
valence states. Spin-orbit interactions are included. We use the cutoff energy of $45$~Ha, and a 12$\times$12$\times$12 
Monkhorst-Pack $\bm{k}$-point grid.

We calculate the temperature dependence of the band gap of PbTe with respect to its LDA value, $\Delta E_G(T)=E_G(T)-E_G(\text{LDA})$, and its temperature derivative $\partial E_G/\partial T$. These temperature changes are added to the band gap values obtained using the HSE03 functional to obtain $E_G(T)$ used in the two-band Kane model.
The effective masses of the renormalized bands near L at a finite temperature are then computed using the two-band Kane model as~\cite{Dresselhaus2008a} 
\begin{equation}
 m^*_d(T)/m^*_d(0~\text{K})=E_G(T)/E_G(0~\text{K}),
\end{equation}
where $d$ denotes the parallel or perpendicular direction. We note that due to the small calculated value of the zero-point renormalization of the gap ($\sim 20$ meV)~\cite{Querales2019}, we approximate the values of the effective masses at 0 K with their values obtained using the HSE03 functional.

\subsection{Thermoelectric properties}
To study thermoelectric transport in n-type PbTe, we use the Boltzmann transport theory 
within the
relaxation time approximation. Thermoelectric transport properties 
can be calculated as
\begin{equation}
  \label{eq:transport}
  \begin{split}
    \sigma^{ij}     & = L^{ij}_0 , \\
    S^{ij}          & = - L^{ij}_1 / (eT L^{ij}_0) , \\
    \kappa^{ij}_0   & = L^{ij}_2 / (e^2 T) ,
  \end{split}
\end{equation}
where $\sigma$ is the dc electrical conductivity tensor, $S$ is the Seebeck coefficient tensor, 
$\kappa_0$ is the thermal conductivity tensor defined when the electric field across the material is 
zero,  
$i$ and $j$ are the Cartesian directions, and $e$ is the electron charge. 
The transport kernel functions for the CB are defined by
\begin{equation}
  \label{eq:kernal}
  L_\alpha^{ ij (e)} = \int_{\rm BZ} \frac{ e^2 d \bm{k}}{4 \pi^3} 
  \left( - \frac{\partial f}{\partial E} \right) \tau_{\bm{k}{,\rm tot}} 
  v^i_{\bm{k}} v^j_{\bm{k}} \left( E_{\bm{k}} - E_F \right)^\alpha,	
\end{equation}
where $E_{\bm{k}}$ and $v^i_{\bm{k}}$ are the energy and the  
group velocity of an electronic state with the crystal momentum  $\bm{k}$, 
$f$ is the equilibrium Fermi-Dirac occupation function, $E_F$ is the Fermi level, and  
$\tau_{\bm{k}{,\rm tot}}$ is the relaxation time.
Unlike $\kappa_0$, the total thermal conductivity $\kappa$ is defined at zero electric 
current across the material, and is the sum of the lattice contribution $\kappa_L$ and the electronic
contribution $\kappa_e$, which can be given as
\begin{equation}
  \label{eq:kappa}
  \kappa = \kappa_L + \kappa_e=\kappa_L+\kappa_0 - T \sigma S^2.
\end{equation}
Since PbTe is cubic, $\sigma^{ij}$, $S^{ij}$, and
$\kappa^{ij}_0$ can be expressed as
\begin{equation}
    \sigma^{ij} = \delta_{ij} \sigma; \>\>
    S^{ij} = \delta_{ij} S; \>\>
    \kappa^{ij}_0 = \delta_{ij} \kappa_0; 
\end{equation}
where $\delta_{ij}$ is the Kronecker delta, and $\sigma$, $S$, and $\kappa_0$ are given as: $\sigma = \sum_i \sigma^{ii} /3$, $S = \sum_i S^{ii} /3$, and $\kappa_0 = \sum_i \kappa_0^{ii} /3$. 

We note that the superscript index $(e)$ in Eq.~\eqref{eq:kernal} refers to electrons. 
We also include the contribution from holes by treating the VB as a mirror-image
of the CB in the two-band Kane model. The analogue definitions of the transport kernel functions for holes 
are obtained by simply substituting $E_F$ by $(-E_G-E_F)$ in 
Eq.~\eqref{eq:kernal}.
By combining the contributions from electrons and holes, we obtain the total transport 
kernel functions 
$L_\alpha = L_\alpha^{(e)} + L_\alpha^{(h)}$.
Our calculations show that the effect of holes on the electronic transport properties is negligible
at the optimum doping concentrations for TE applications ($\sim$10$^{19}$~cm$^{-3}$).

The total relaxation time
$\tau_{\bm{k}{,\rm tot}}$ is determined by the contributions of 
different scattering mechanisms: acoustic (ac) phonons, transverse optical 
(TO) phonons, longitudinal optical (LO) phonons, and ionized 
impurities (imp)~\cite{Song2017,Cao2018,Ravich1971b,Zayachuk1997}, and can be calculated via Matthiessen's rule:
\begin{equation}
  \tau^{-1}_{\bm{k}{,\rm tot}} = \tau^{-1}_{\bm{k}{,\rm ac}} 
  +\tau^{-1}_{\bm{k}{,\rm LO}}+\tau^{-1}_{\bm{k}{,\rm TO}} 
  +\tau^{-1}_{\bm{k}{,\rm imp}}.
  \label{eq:tautot}
\end{equation}
The relaxation time of a single scattering channel
is given by~\cite{Sohier2014,Gunst2016}
\begin{equation}
  \label{eq:tauk}
  \tau^{-1}_{\bm{k}} = \frac{2 \pi}{\hbar} \sum_{\bm{q} } 	
  \frac{1-f_{\bm{k}'}}{1-f_{\bm{k}}}
  \left(1 - \hat{\bm{v}}_{\bm{k}} \cdot \hat{\bm{v}}_{\bm{k}'} \right)
  S_{{\bm{k}}}^{\bm{k}' },
\end{equation}
where $S_{{\bm{k}}}^{\bm{k} ' }$ denotes the transition rate 
from initial $\bm{k}$ to final state $\bm{k} '$ due to  
scattering, and $\hat{\bm{v}}_{\bm{k}}$ is the unit vector in the direction of the group velocity at  $\bm{k}$. 
The velocity factor 
$\left(1 - \hat{\bm{v}}_{\bm{k}} \cdot \hat{\bm{v}}_{\bm{k}'} \right)$
accounts for the change of direction of scattered carriers.
For electron-phonon scattering, we can write
\begin{equation}
  \label{eq:skkq}
  \begin{split}
  S_{{\bm{k}}}^{\bm{k} +\bm{q} }=  \abs{g_{{\bm{k}}}^{\bm{k} +\bm{q} } }^2 &
  \{  
     N^0(\omega_{\bm{q}})  
    \delta(E_{\bm{k}} + \hbar \omega_{\bm{q}} - E_{\bm{k}+\bm{q}})  \\
    & +[ N^0(\omega_{\bm{q}})  +1 ]\delta(E_{\bm{k}} - \hbar \omega_{\bm{q}} - E_{\bm{k}+\bm{q}})
  \}   ,    
  \end{split}
\end{equation}
where $N^0$ and $\omega_{\bm{q}}$ are the equilibrium distribution and the 
frequency of a phonon with the crystal momentum $\bm{q}$, 
and $g_{{\bm{k}}}^{\bm{k} +\bm{q} }$ is the electron-phonon 
matrix element. The two terms in the curly brackets correspond 
to phonon absorption and emission, respectively.

We parametrize the electron-phonon matrix elements  $g_{{\bm{k}}}^{\bm{k} +\bm{q} }$ due to acoustic, TO and LO phonons as described in detail in our previous work~\cite{Cao2018}. This requires calculations of acoustic and optical deformation potentials, phonon frequencies, elastic and dielectric constants~\cite{Cao2018}. Some of these parameters, such as phonon frequencies, elastic and dielectric constants, can be computed straightforwardly from first principles, using DFPT. Our earlier work~\cite{Murphy2018} describes several methods to calculate acoustic deformation potentials of PbTe from first principles, one of which uses DFPT. We use the same DFPT method to obtain optical deformation potentials of PbTe. The values of all these parameters are listed in Table I of Ref.~\cite{Cao2018}. Since the VB is described as a mirror image of the CB, the absolute values of deformation potentials for the VB of PbTe are taken to be the same as those for the CB. All these parameters are calculated using the LDA excluding SOC that gives a positive band gap and the correct character of the conduction and valence band states near the L point in PbTe, in contrast to the LDA including SOC~\cite{Murphy2018}\footnote{We note that we obtain similar values of acoustic deformation potentials using the LDA excluding SOC and the HSE03 functional including SOC~\cite{Murphy2018}.}. We note that we do not include the temperature changes of the parameters characterizing the strength of electron-phonon coupling in this work. We also account for ionized impurity scattering using the Brooks-Herring model with the Thomas-Fermi model for carrier screening, which was also used for screening of LO phonon scattering~\cite{Cao2018}.

\section{Results and Discussion}
\label{sec:results}

\begin{figure}[h!]
  \centering
  \includegraphics{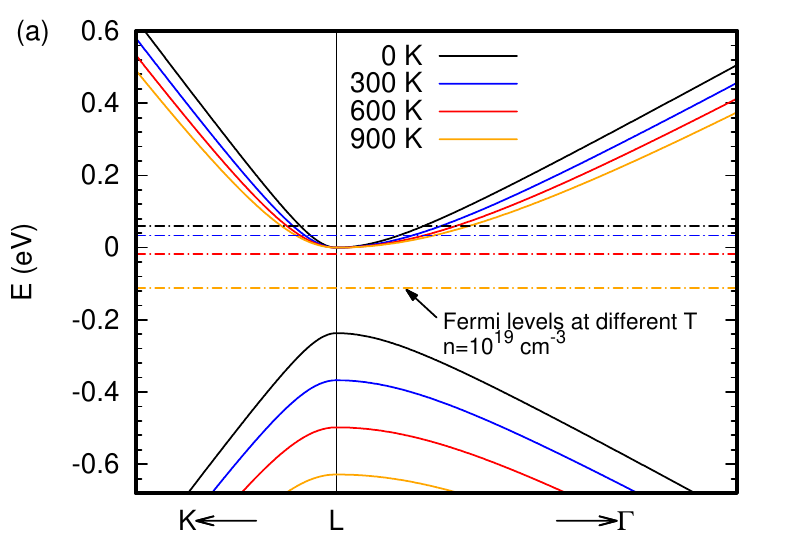}
  \includegraphics{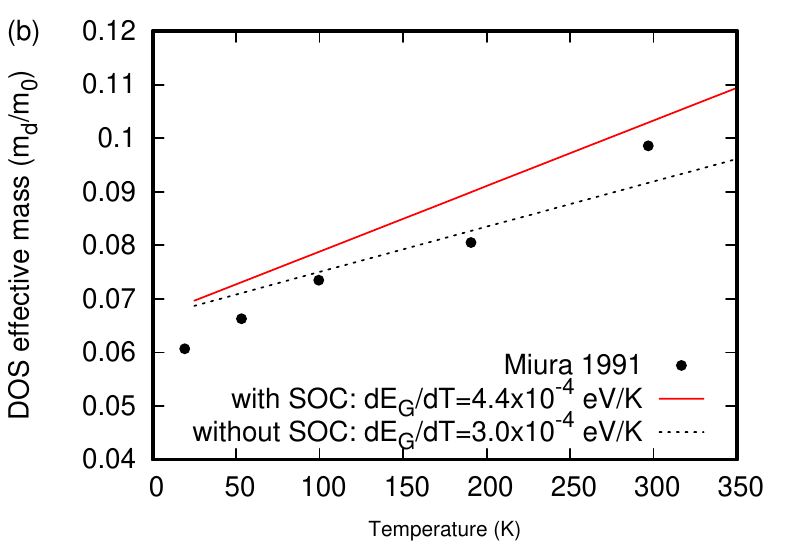}
  \caption{ (a) The lowest conduction band and the highest valence band of PbTe near the L point and the 
    Fermi level for the doping concentration of $n=10^{19}$~cm$^{-3}$ at different temperatures calculated using
     $\partial E_G/\partial T\approx 4.4\times 10^{-4}$ eV/K. 
    The conduction band minima from the different temperature calculations are aligned.
    (b) Temperature dependence of the density-of-states (DOS) effective mass calculated with the 
    two-band Kane model (lines) and measured experimentally (dots)~\cite{Bauer1991}. }
  \label{fig:TdepBS}
\end{figure}

Using the HSE03 hybrid functional, we obtain the band gap value of $E_G=0.237$~eV for PbTe,~\cite{Murphy2018}, which compares 
well to the experimental value of 0.19~eV at 4 K~\cite{DALVEN1974}. 
We calculate the temperature dependence of the band gap using DFPT and the LDA excluding and including SOC.
The top valence and bottom conduction bands at L of PbTe correspond to the representations L$^{6+}$ and L$^{6-}$, respectively~\cite{Kohn1973}, but that order is inverted in the LDA including SOC~\cite{Murphy2018}. To account for the correct ordering of these states, we define the direct gap at L as $E_G = E_{\text{L}^{6-}}-E_{\text{L}^{6+}}$~\cite{Querales2019}.
Using a linear fit for $E_G$ with respect to $T$ in the range of $200\--800$ K, we compute $\partial E_G/\partial T \approx 3.0 \times 10^{-4}$ eV/K and $\partial E_G/\partial T\approx 4.4\times 10^{-4}$ eV/K excluding and including SOC, respectively~\cite{Querales2019}. Both values compare very well to the available optical absorption measurements of $\partial E_G / \partial T\sim 3.0-5.1 \times 10^{-4}$ eV/K~\cite{Snyder2013,Gibson1952,Tauber1966,Saakyan1966,Baleva1990}. These results suggest that accounting for SOC or the correct order of the states near the gap does not affect the computed $\partial E_G/\partial T$ values very much.

The temperature dependence of the CB and VB near the L point is shown in 
Fig.~\ref{fig:TdepBS}(a), where the CB minima from different
temperature calculations are aligned.
The increasing effective mass with temperature obtained in our calculations has also been observed experimentally. 
As shown in Fig.~\ref{fig:TdepBS}(b),
the computed conduction band density-of-states (DOS) effective mass  
is in fairly good agreement with the measurements of Ref.~\cite{Bauer1991}, particularly when we use the value of $\partial E_G/\partial T\approx 4.4\times 10^{-4}$ eV/K which includes the effects of SOC. Therefore, $\partial E_G/\partial T\approx 4.4\times 10^{-4}$ eV/K is used in the rest of the paper. We have verified that the conclusions of this work do not change if we use the value of $\partial E_G/\partial T\approx 3.0\times 10^{-4}$ eV/K in our simulations. 

\begin{figure}[h]
  \centering
  \includegraphics{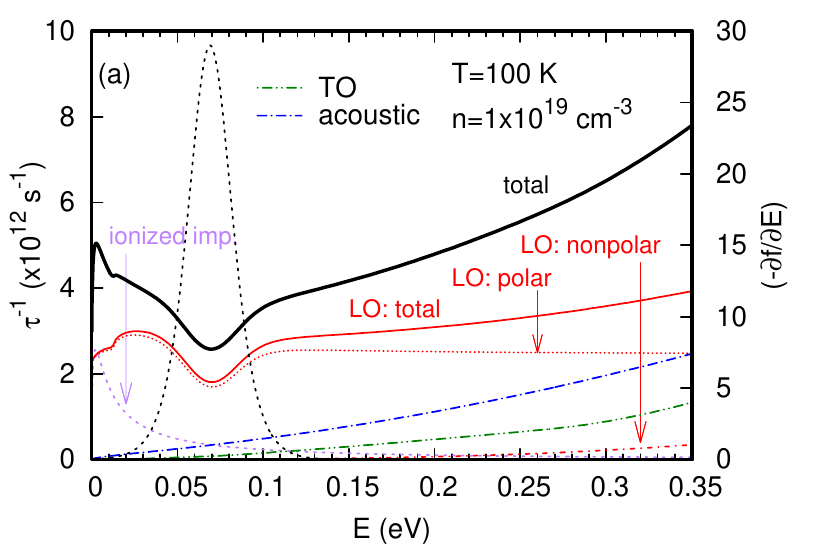}
  \includegraphics{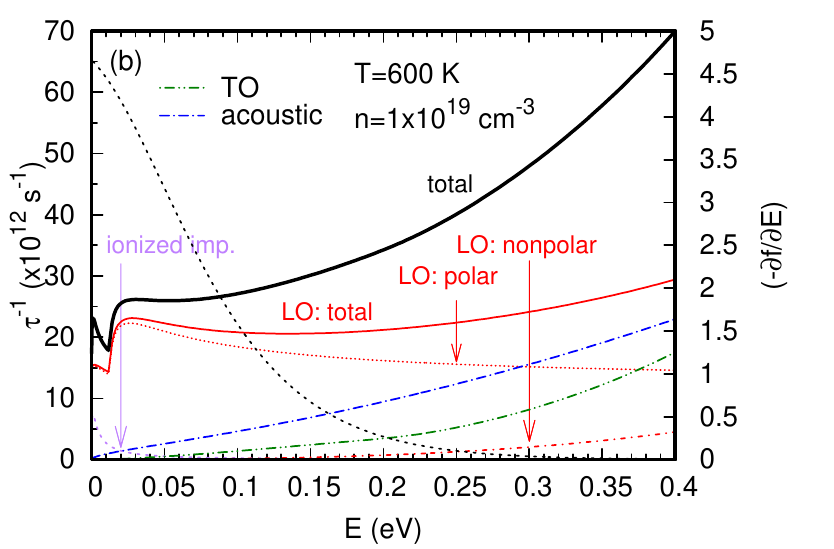}
    
  \caption{Calculated energy dependence of the inverse of the 
    relaxation time ($\tau^{-1}$) for longitudinal optical (LO), 
    transverse optical (TO) and acoustic phonon scattering, and ionized 
    impurity scattering in PbTe for the doping concentration of $n=1\times 
    10^{19}$~cm$^{-3}$ at: (a) 100~K, and (b) 600~K. Black dashed curves represents the derivative of the Fermi-Dirac 
    distribution function with respect to energy, indicating the energy 
    region contributing to electronic transport at different temperatures. }
  \label{fig:invRT}
\end{figure}

We next show the calculated energy dependence 
of the inverse relaxation time associated with the 
four scattering mechanisms ($\tau^{-1}_{{\rm ac}}$, 
$\tau^{-1}_{{\rm LO}}$, $\tau^{-1}_{{\rm TO}} $, 
$\tau^{-1}_{{\rm imp}}$) and $\tau^{-1}_{{\rm tot}} $
for $T=100~K$ and $T=600~K$ at the doping 
concentrations of $n=1\times 10^{19}$~cm$^{-3}$, see Fig.~\ref{fig:invRT}.
We also plotted $(- \partial f/\partial E)$ to indicate the energy 
range that contributes to electronic transport and understand the relative 
importance of different scattering mechanisms in this range.
We find that LO phonon scattering is the strongest scattering mechanism, while acoustic and, to a lesser extent,
TO scattering become comparable only when $T$ and doping are very high.
LO scattering includes long-range polar and
short-range non-polar interactions, whose relaxation times are also plotted 
in Fig.~\ref{fig:invRT}.
Since the matrix elements $g_{{\bm{k}}}^{\bm{k} +\bm{q} } $ of these two contributions are added up,
the non-polar
contribution considerably modifies the energy dependence of $\tau_{{\rm LO}}$ despite
its relatively small magnitude.
In general, ionized impurity scattering is negligible compared to 
electron-phonon scattering, except at
low $T$ and low doping concentrations.
At low temperatures ($T$=100~K), there is a dip in $\tau^{-1}_{{\rm LO}}$ around the Fermi level, which is 
a consequence of the Pauli exclusion principle for this inelastic process (see~\ref{appx0}).

\begin{figure}[h]
  \centering
  \includegraphics{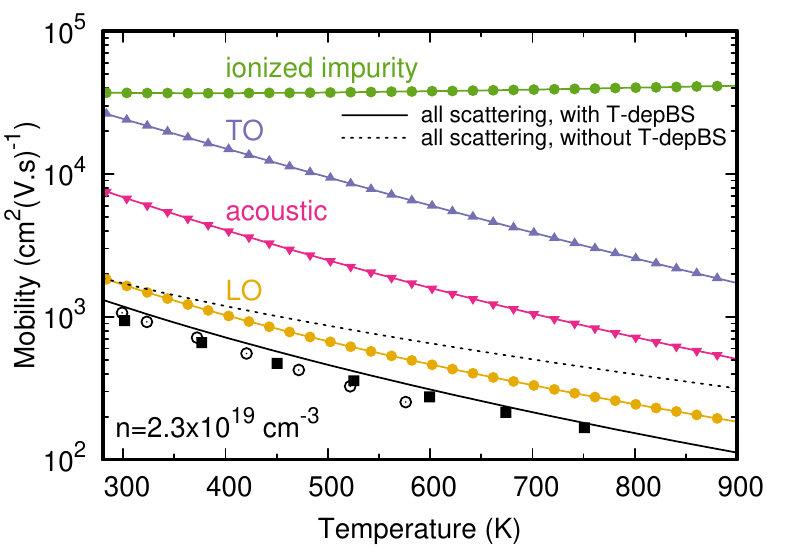}
  \caption{Electronic drift mobility versus temperature for the doping 
    concentration of $n=2.3\times 10^{19}$~cm$^{-3}$. Solid black line is 
    calculated including all
    scattering channels, while
    colored lines correspond to the contributions to the mobility from individual
    scattering channels. 
    Dashed black line represents the total mobility calculated without
    accounting for the temperature dependence of the electronic band structure (T-depBS).
    Experimental data are shown in full squares (Ref.~\cite{Snyder2014}) 
    and circles (Ref.~\cite{Ravich1971b}). }
  \label{fig:mobiT}
\end{figure}

The importance of accounting for the temperature dependence of the band structure (T-depBS)
to compute the electronic transport properties of PbTe is illustrated in Fig.~\ref{fig:mobiT}. The electronic mobility for $n$=$2.3\times 10^{19}$~cm$^{-3}$ as a function of temperature excluding and including the T-depBS is given by the black solid and dashed lines, respectively. Squares and circles show the mobility measurements from Refs.~\cite{Ravich1971b,Snyder2014}.
 Including the T-depBS decreases the mobility about 25\% at 300~K and 50\% at 900~K, and gives a much better agreement with the experimental values and their temperature dependence. The difference between our results including and excluding the T-depBS becomes larger as $T$ increases. This is because $m^*_\parallel$ and $m^*_\perp$ increase with $T$ due to the increased band gap, leading to a decrease of the band curvature near the band edge and an increase of the radius of constant energy surface. These effects result in smaller group velocities and a larger phase space for scattering, thus reducing the mobility.  

We also calculated the individual contributions of various scattering channels to the electronic mobility of PbTe, 
shown by the colored lines in Fig.~\ref{fig:mobiT}. LO phonon scattering is the dominant scattering channel limiting the mobility between 300~K and 900~K, even at high doping concentrations, since screening  is relatively weak due to a large dielectric constant of PbTe~\cite{Cao2018}, and also does not affect the non-polar contribution. In contrast, TO phonon scattering is the weakest electron-phonon scattering mechanism even for high temperatures, since this type of scattering between the CB minima at L and the zone center TO mode is forbidden by symmetry~\cite{Cao2018}. Acoustic phonon scattering is the second strongest scattering channel, whose importance increases at high temperatures and doping concentrations. 

 We note that the T-depBS makes the mobility limited 
 by individual electron-phonon scattering channels exhibit very similar
 $T$ dependence, as shown by the colored lines in Fig.~\ref{fig:mobiT}. 
 One would expect that LO phonon
 scattering has a characteristic $T$ dependence that is different
 from that of acoustic phonons due to the different energy dependence of their
 relaxation times $\tau$. However, the $T$ dependence
 of the mobility is not very sensitive to the energy dependence of 
 $\tau$ over a large range of $T$ when the temperature dependence of
 the band structure is accounted for in our calculations.
 Instead, the $T$ dependence of the mobility comes from three sources: 
 (1) the $T$ dependence of phonon occupations 
 ($N^0 \approx k_B T/\hbar \omega$ for $k_B T \gg \hbar \omega$),
 (2) the $T$ dependence of the electronic band structure,
 and (3) the $T$ dependence of the Fermi level $E_F$.
 Our results suggest that it is not reliable to deduce the relative weight of 
 different electron-phonon scattering processes only by considering the $T$
 dependence of the mobility in the case of direct narrow-gap semiconductors with
 band gaps strongly renormalized by temperature.

\begin{figure}[h]
  \centering
  \includegraphics{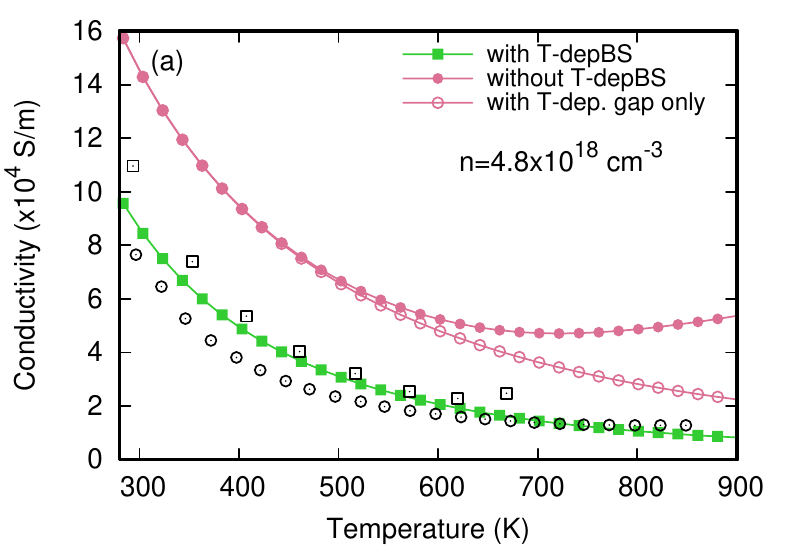}   
  \includegraphics{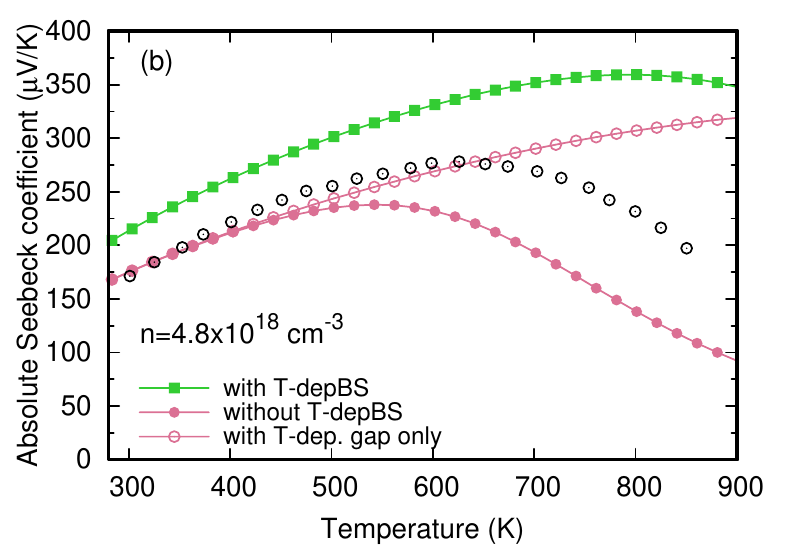} 
  \caption{ Calculated and experimental temperature dependence of (a) the conductivity 
    and (b) the absolute Seebeck coefficient of n-type PbTe. 
    Experimental data are shown in circles (Ref.~\cite{Snyder2014}) and 
    squares (Ref.~\cite{Vineis2008}). T-depBS in the legend stands for the temperature dependence of the band structure, while T-dep. gap only corresponds to the temperature dependence of the band gap only.}    
  \label{fig:vsTwwoTdep}
\end{figure}

The calculated temperature dependence of the electrical conductivity and the Seebeck 
coefficient for $n$=$4.8\times 10^{18}$~cm$^{3}$ including and excluding the 
T-depBS is illustrated in Fig.~\ref{fig:vsTwwoTdep}. 
At high $T$, the discrepancy between the results excluding and including the 
T-depBS comes from two factors:
(1) increasing $E_G$ with $T$ which reduces the hole contribution to electronic transport,
and (2) increasing effective masses with $T$.
To identify the role of each of these factors, we calculate the conductivity and Seebeck 
coefficient using the temperature-dependent gap values but keeping the effective masses constant at their 0~K values,
shown by the red lines with empty circles in Fig.~\ref{fig:vsTwwoTdep}. 
The difference between the red empty and full circles
shows the effect of increased hole concentration due to the increased $E_G$:
(1) the electrical conductivity increases, 
and (2) the Seebeck coefficient decreases because the energy 
term $(E_{\bm{k}} - E_F)$ in Eq.~\ref{eq:kernal} is negative for $S^{\rm (h)}$.
These effects become strong only at low doping concentrations and high $T$. 
On the other hand, as shown by the difference
between the green full squares and the red empty circles, the temperature induced band flattening
increases the Seebeck coefficient and decreases the conductivity over the entire temperature range.
We note that these effects
become weaker as the doping concentration increases because the Fermi level increases and the electronic states relevant for transport are less influenced by the temperature variations of effective masses.

\begin{figure*}[h!]
  \centering
  \includegraphics{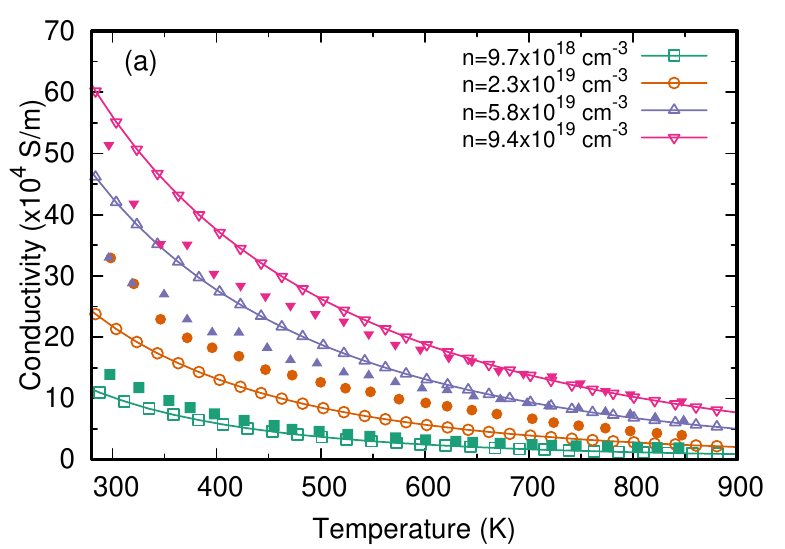}
  \includegraphics{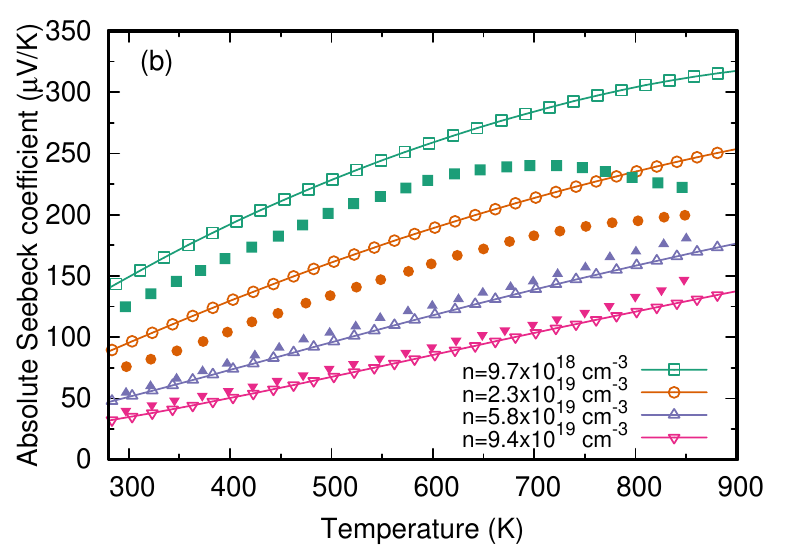} \\  
  \includegraphics{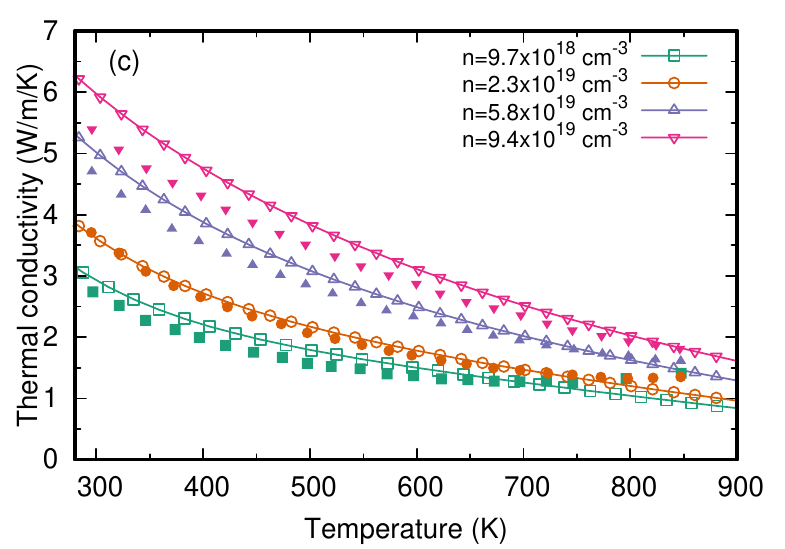}
  \includegraphics{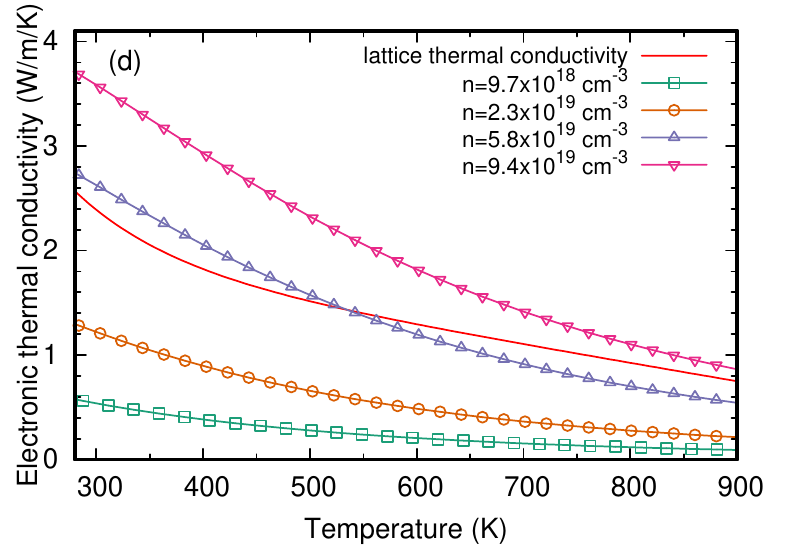}
  \caption{Calculated and experimental temperature dependence of (a) the electrical conductivity,
    (b) the absolute Seebeck coefficient, (c) the total thermal conductivity and
    (d) the electrical thermal conductivity of n-type PbTe for several doping concentrations. 
    Experimental data are shown in full symbols (Ref.~\cite{Snyder2014}).}
  \label{fig:vsT}
\end{figure*}

The comparison between the computed and measured electrical conductivity as a function of temperature
for $n$=$4.8\times 10^{18}$~cm$^{3}$ clearly shows that it is important to include the T-depBS  
to obtain good agreement with experiments~\cite{Snyder2014,Vineis2008}
(see Fig.~\ref{fig:vsTwwoTdep}(a)). On the other hand, the Seebeck coefficient is somewhat overestimated
compared to experiment when the T-depBS is accounted for  (Fig.~\ref{fig:vsTwwoTdep}(b)),
likely due to the fact that our first principles calculations somewhat
overestimate the effective masses of PbTe (Fig.~\ref{fig:TdepBS}(b)).  
We also note that the calculated Seebeck coefficient for $n=4.8\times 10^{18}$~cm$^{-3}$ keeps increasing with temperature, and is overestimated at $T>$700~K  (see Fig.~\ref{fig:vsT}). In PbTe, the valence bands at $\Sigma$ become aligned with the valence bands at L for $T\sim 620$~K~\cite{Snyder2013,Querales2019}. This effect will also increase the hole concentration and decrease the Seebeck coefficient with respect to our present results.

The temperature dependence of the electrical conductivity, Seebeck coefficient, total
and electronic contribution to the thermal conductivity 
for different doping concentrations ($n$=$9.7\times 10^{18}$, $2.3\times 10^{19}$, 
$5.8\times 10^{19}$, and $9.4\times 10^{19}$~cm$^{-3}$)
calculated including the T-depBS is plotted in Fig.~\ref{fig:vsT}.
The lines with empty symbols represent our calculations, while the full symbols
show the measurements from Ref.~\cite{Snyder2014}. The lattice thermal 
conductivity values are taken from our previous first principles calculations~\cite{Murphy2016}.
For the doping concentrations above 
$5.8\times 10^{19}$~cm$^{-3}$, the electronic contribution to the thermal 
conductivity becomes higher than the lattice contribution.
The computed TE transport properties are generally in good agreement with
experiments for a wide range of doping concentrations and temperatures.

The TE transport properties of n-type PbTe as a function of the doping concentration with and without the T-depBS at $T$=300~K are given in \ref{appx1}. We find that the calculated transport properties accounting for the T-depBS are all in very good agreement with the measurements from several experiments~\cite{Ravich1971b,Harman1996,Snyder2014,Vineis2008}. Including the temperature dependence of the band structure in the first principles calculations of the thermoelectric transport properties is thus critical for accurately reproducing their dependence on the temperature and doping concentration.  

\begin{figure*}[h!]
  \centering
  \includegraphics[height=4.57cm]{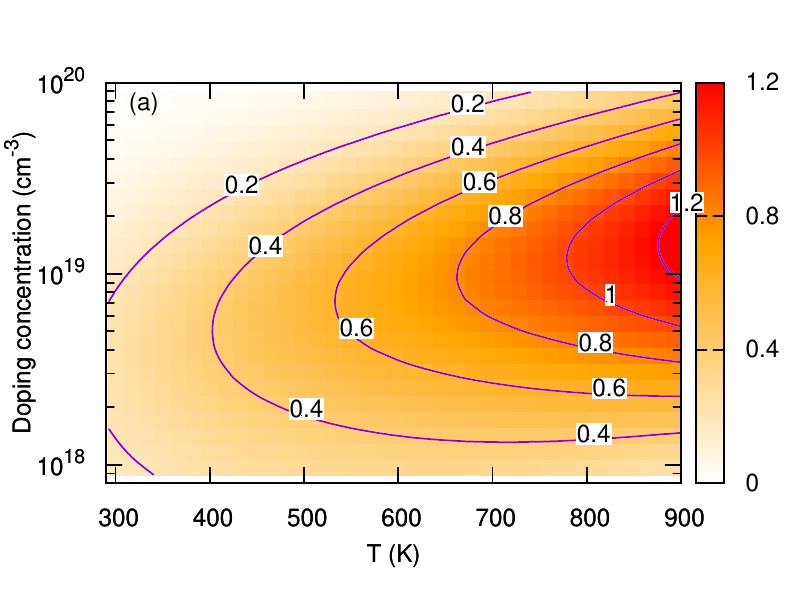}
  \includegraphics[height=4.57cm]{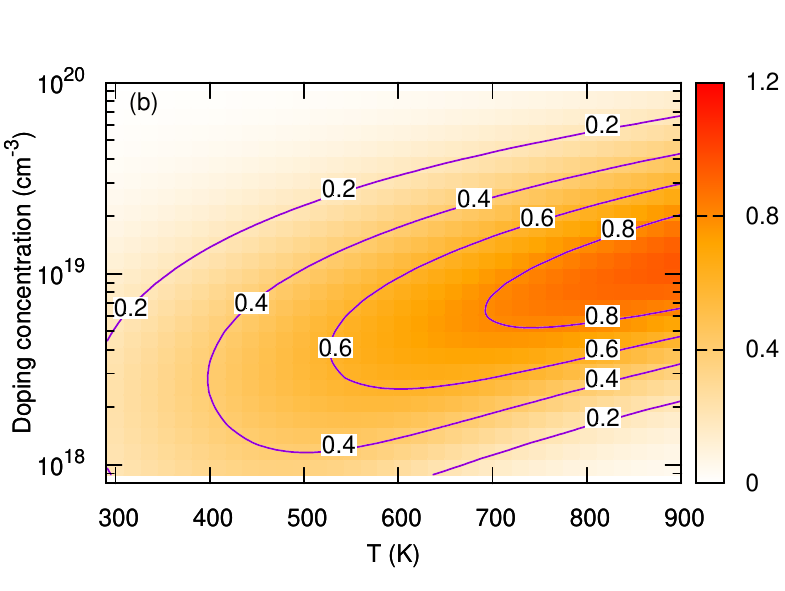}
  \includegraphics[height=4.18cm]{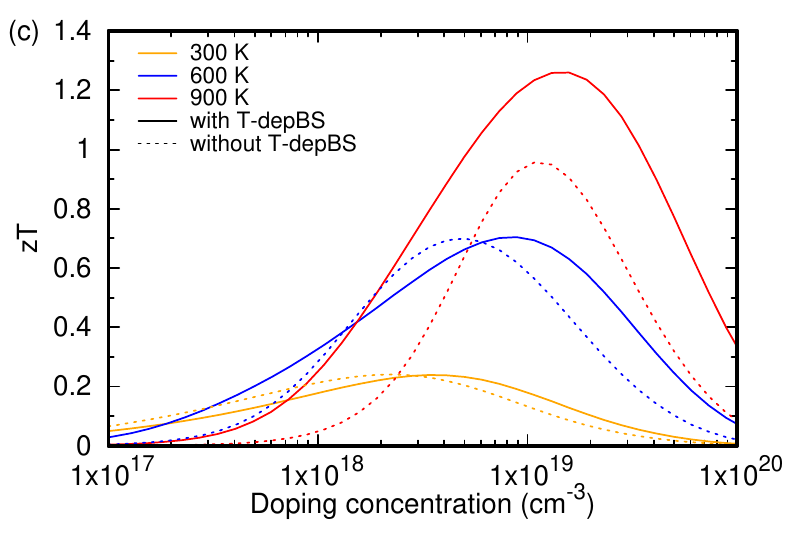}
  \caption{Color maps of the calculated figure of merit $zT$ versus 
    temperature and doping concentration (a) including and (b) excluding the temperature dependence of the electronic band structure (T-depBS). (c) Calculated $zT$ versus doping concentration for different temperatures including and excluding T-depBS.}
  \label{fig:zT}
\end{figure*}

We also calculate the $zT$ of n-type PbTe including and excluding the T-depBS versus $T$ and doping concentration,
as shown by the color maps in Fig.~\ref{fig:zT}(a) and (b). The color bar indicates the values of $zT$, while the contour
lines trace the iso-values of $zT$. For a more quantitative representation, Fig.~\ref{fig:zT}(c) displays the $zT$ dependence on the doping concentration at $T$=300, 600, and 900~K. The increasing band gap with $T$ in PbTe leads to: (1) the higher peak $zT$ value for $T$ above 650~K, and (2) the smaller doping
 concentration dependence of $zT$ around its peak, which leads to the higher average $zT$ over a range of 
 temperatures. 

\begin{figure}[h!]
  \centering
  \includegraphics[width=0.9\columnwidth]{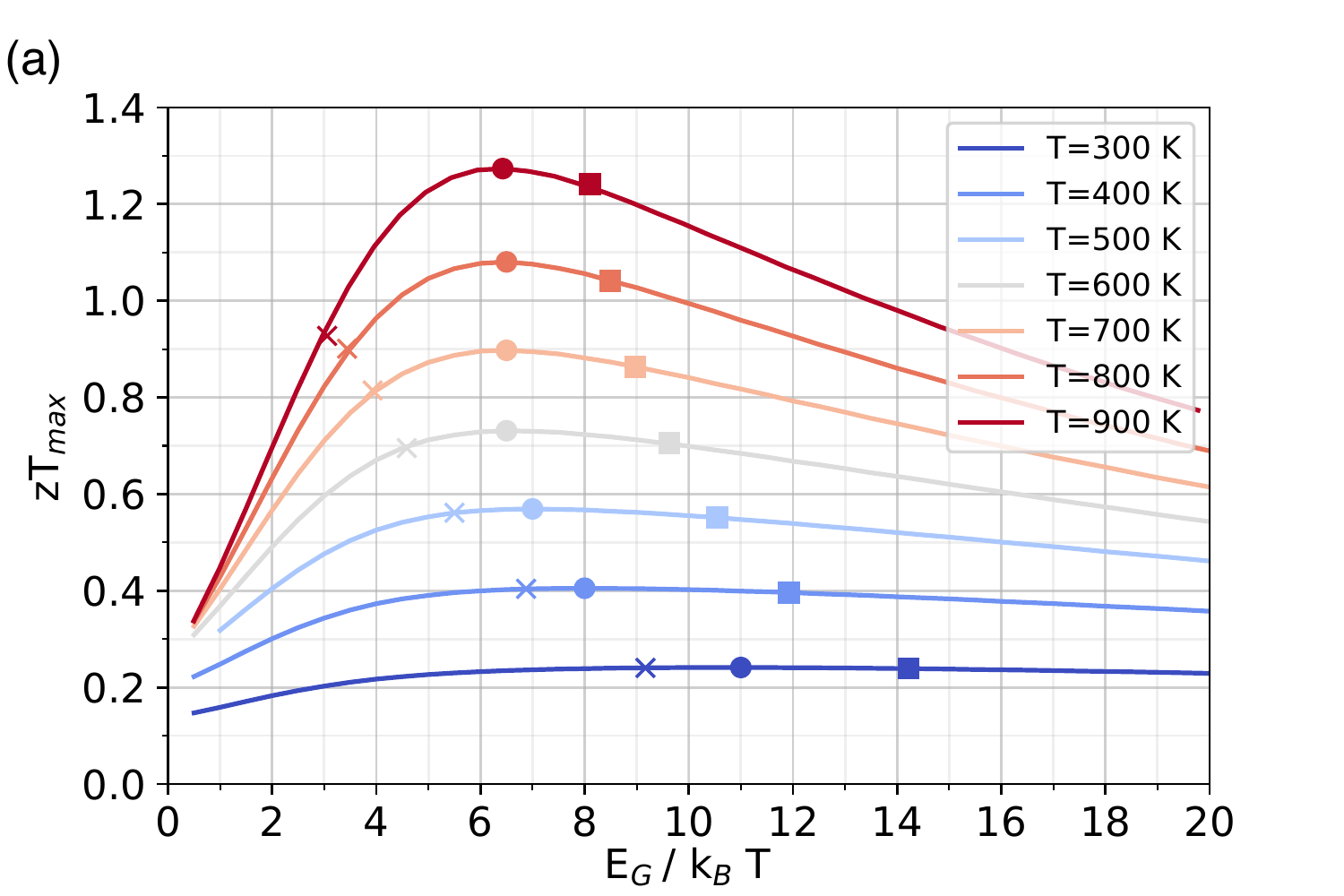} \\
  \includegraphics[width=0.9\columnwidth]{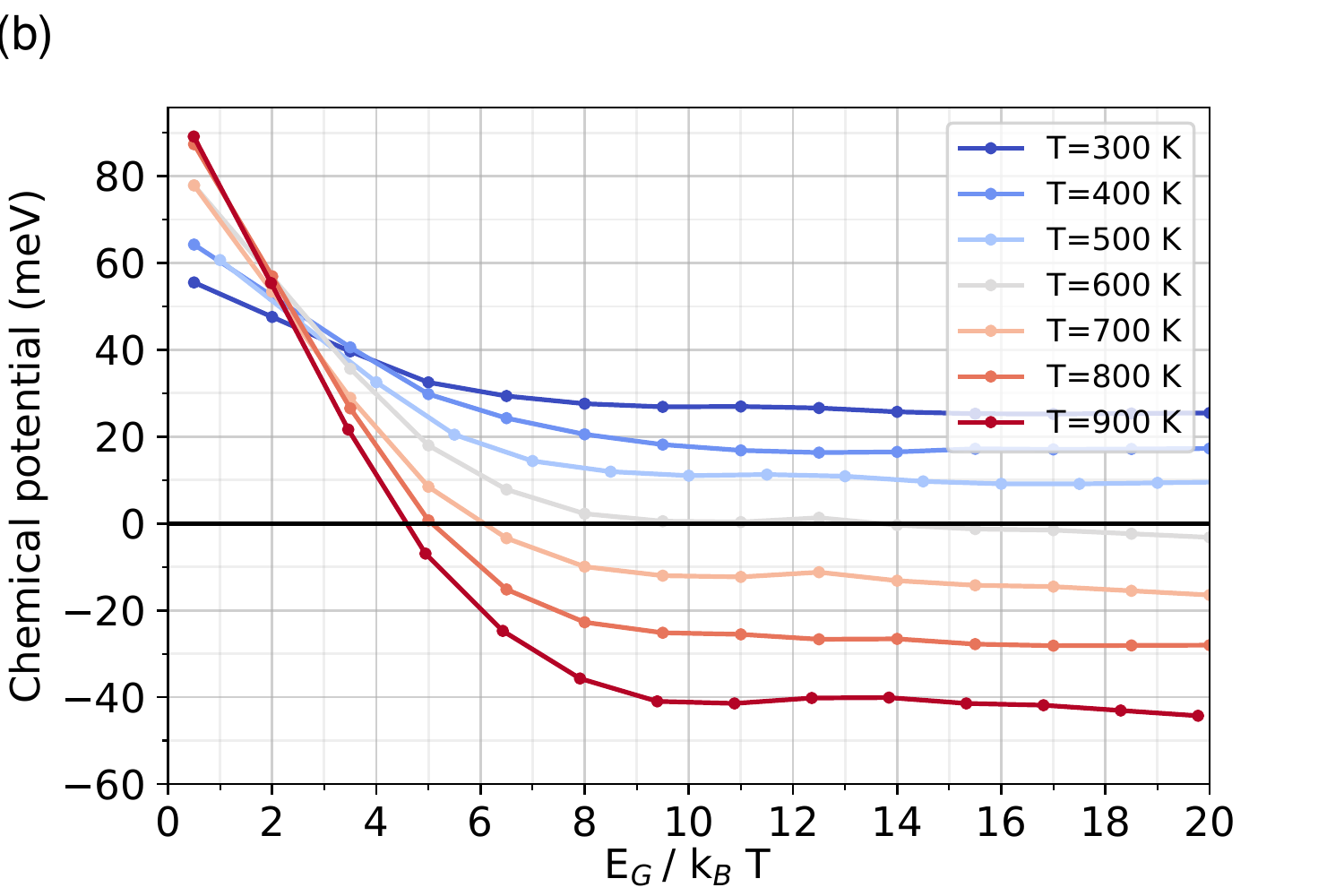}
  \caption{(a) Maximum value of $zT$ ($zT_{\rm max}$) and
    (b) chemical potential corresponding to $zT_{\rm max}$ as a function
    of the band gap $E_G$ in units of thermal energy $k_B T$ for different
    temperatures between 300 and 900~K.
    $zT_{\rm max}$ values are obtained by adjusting the doping concentration for
    each $E_G$ and $T$. The band gap values giving maximum $zT_{\rm max}$ values are
    indicated by circles.
    The calculated band gap values of PbTe where their temperature variation is accounted for are indicated by squares, while the ground state band gap value of $E_G=0.237$~eV is given by crosses. 
  }
  \label{fig:maxzT}
\end{figure}

To understand why the T-depBS enhances the $zT$ values of n-type PbTe, we perform a thought experiment where we vary the band gap values in the two-band Kane model independently from temperature. For each value of the gap, we also change the effective masses according to the two-band Kane model. All other parameters of our model are kept fixed. For each $T$ and band gap value, we calculate the maximal value of $zT$, $zT_{\rm max}$, by optimizing the doping concentration. Figs.~\ref{fig:maxzT}(a) and (b) show $zT_{\rm max}$ and the optimal Fermi level as a function of the band gap for $T$ ranging from 300~K to 900~K. The optimal values of the gap for which $zT_{\rm max}$ becomes maximal are given by the circles in Fig.~\ref{fig:maxzT}(a). For high temperatures, $zT_{\rm max}$ reaches a peak value at $E_G \approx 6.5 k_B T$ and then decreases with increasing band gap. For low temperatures,  $zT_{\rm max}$ is constant for large enough gaps and does not show a clear peak. These values and trends are in agreement with the previous work of Sofo and Mahan~\cite{Sofo1994}.

\begin{figure}[h!]
\centering
  \includegraphics[width=0.8\columnwidth]{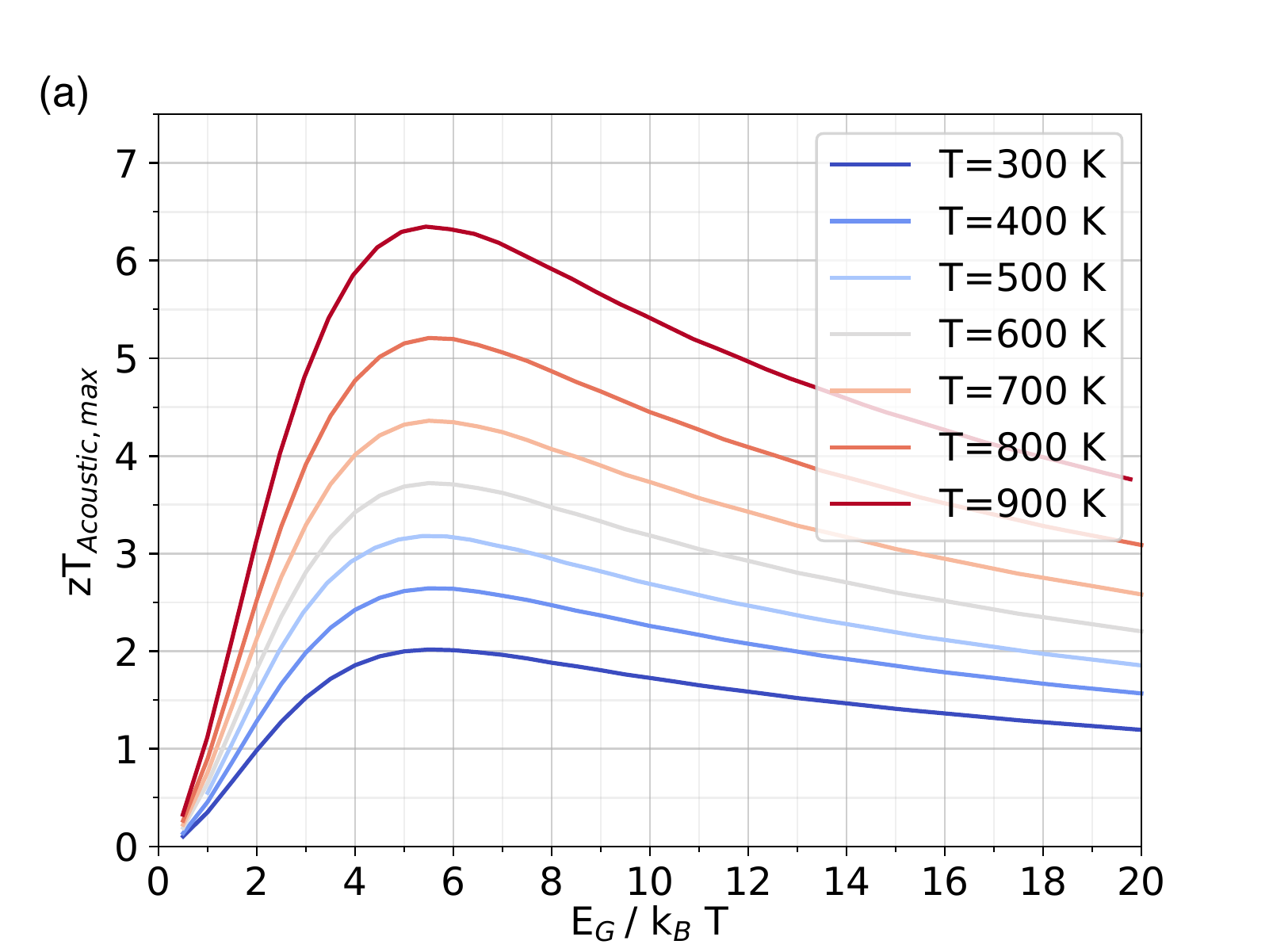}
  \includegraphics[width=0.8\columnwidth]{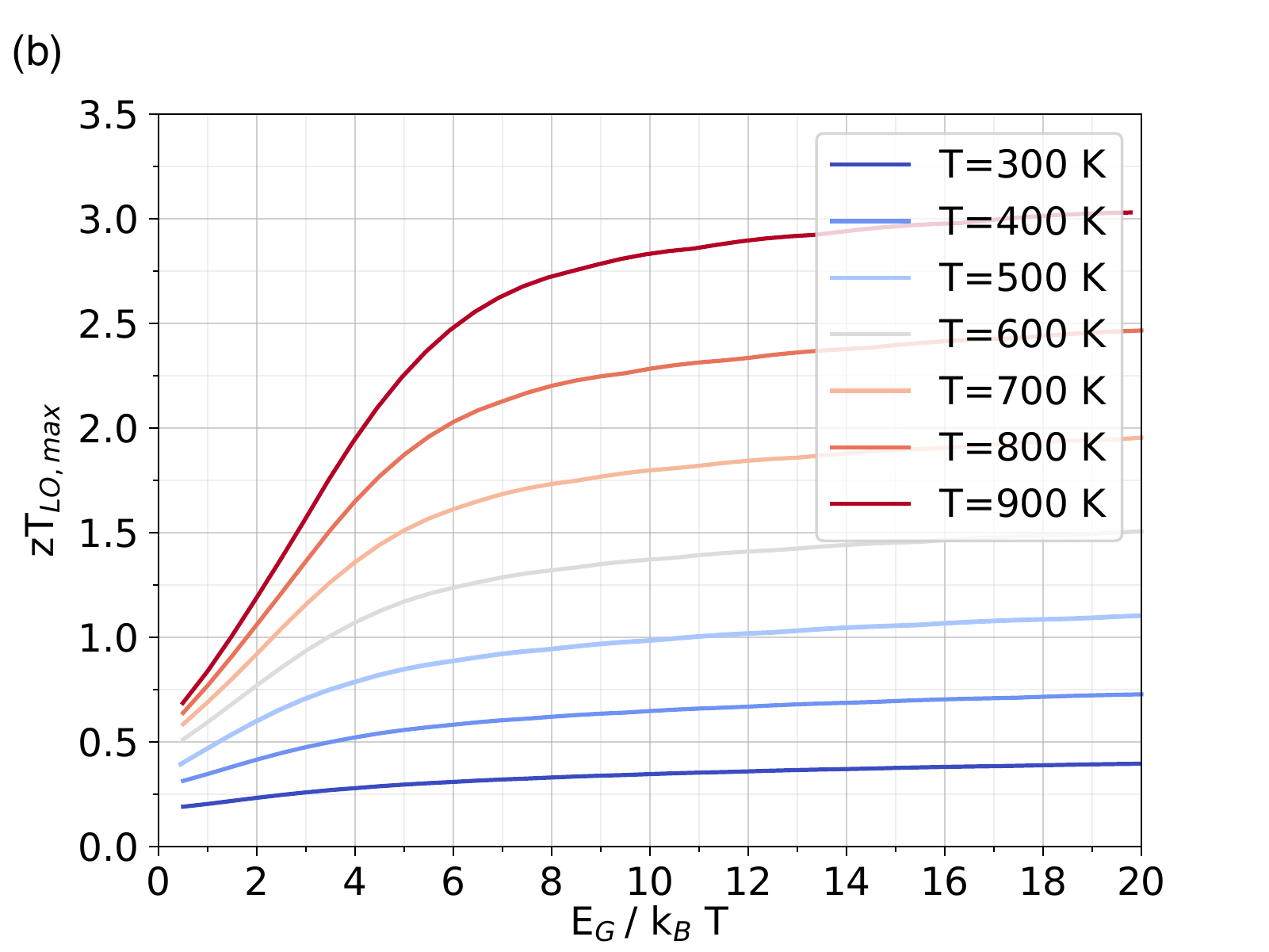}
  \caption{Maximum value of $zT$  ($zT_{\rm max}$) as function of the band gap $E_G$ in units of thermal
    energy $k_B T$ due to: (a) acoustic phonon scattering, and (b) polar longitudinal phonon
    scattering. In these plots, only the $zT_{\rm max}$ dependence on $E_G$ is relevant, while the $zT$ values on the vertical axis are given on an arbitrary scale.
  }
  \label{fig:maxzTscat}
\end{figure}

The observed decrease of $zT_{\rm max}$ with decreasing band gap for $E_g<6k_BT$ is due to the appearance of minority carriers, which strongly reduce the Seebeck coefficient and $zT$. For $E_G> 6k_BT$, $zT_{\rm max}$ can have different trends depending on the dominant scattering mechanism~\cite{Sofo1994}. To gain a better insight into this,
we calculate $zT_{\rm max}$ due to polar LO and acoustic phonon scattering, plotted in Figs.~\ref{fig:maxzTscat}(a) 
and (b), respectively. Such $zT$ values are given on an arbitrary scale,
since we are only interested in the $zT_{\rm max}$ trends.
For acoustic scattering, $zT_{\rm max}$ always peaks at $E_G = 5.5 k_B T$ at different temperatures. 
For polar LO scattering, $zT_{\rm max}$ rises slowly with increasing $E_G$.
These effects are related to the dependence of the quality factor $B$,
which was first introduced by Chasmar and Stratton~\cite{CHASMAR1959}, on the effective masses, which change
with temperature in direct gap semiconductors~\cite{Sofo1994}.
In the case of PbTe, polar LO phonon scattering has a major effect at low temperatures and low doping concentrations, which is why the $zT_{\rm max}$ curves in Fig.~\ref{fig:maxzTscat}(a) are nearly flat around room temperature.  
When $T$ increases and screening becomes larger at high doping concentrations, 
acoustic and non-polar optical phonon scattering contributions exceed that of polar LO phonons, leading to  
the $zT_{\rm max}$ peaks in the range of 6 - 10$ k_B T$. 

Finally, we analyze where the actual values of the band gap of PbTe lie on the $zT_{\rm max}$ curves given in Fig.~\ref{fig:maxzT}(a). The squares on the $zT_{\rm max}$ curves correspond to our calculated band gaps renormalized by temperature, while the crosses correspond to the DFT-HSE03 gap value of $E_G=0.237$~eV that does not change with $T$. The $zT_{\rm max}$ values obtained using the gap values that account for the T-depBS are very close to the $zT_{\rm max}$ maxima, especially at high temperatures, as a result of the considerable thermally induced increase of the band gap\footnote{The calculated band gap values are even closer to the $zT_{\rm max}$ maxima at higher temperatures if we use the value of $\partial E_G/\partial T\approx 3.0\times 10^{-4}$ eV/K}. In contrast, if the band gap is constant (or decreasing with $T$), it becomes much smaller than the optimum band gap at high temperatures, leading to a substantial decrease of $zT_{\rm max}$. We thus conclude that the unusual increase of the band gap with temperature is an important factor for the high $zT$ of n-type PbTe. Our results also indicate that other direct narrow-gap semiconductors with positive temperature coefficients resulting in the band gaps of $\sim$ 6-10 $k_B T$ may be potentially good TE materials for a relatively wide range of temperatures, similarly to PbTe.


\section{Conclusion}

We developed the first-principles thermoelectric transport model that includes the temperature variations of
the electronic band structure, and accurately describes the temperature dependence of all thermoelectric
transport properties
for n-type PbTe between 300~K and 900~K.
We also computed the optimum band gap values for which the $zT$ values of n-type PbTe would be maximized, that vary between $10 k_B T$ at 300~K and $6.5 k_B T$ at 900~K.
We showed that the actual band gap values in PbTe are very close to the predicted optimum values   
in a broad range of temperatures, which contributes largely to the good 
thermoelectric figure of merit of n-type PbTe in addition to its low thermal conductivity. 
We propose that materials with positive temperature coefficients producing the band gaps of $\sim$ 6-10 $k_B T$
in a range of temperatures could be promising candidates in the search for efficient thermoelectric materials.

\section*{Acknowledgement}

We thank Tchavdar Todorov for useful discussions.
This work was supported by Science Foundation Ireland (SFI) under Investigators Programme No. 15/IA/3160 
and Natural Science Foundation of Jiangsu Province No. BK20180456. 
We also acknowledge the Irish Centre for High-End Computing (ICHEC) for the provision of computational facilities. 

\appendix

\section{Inelastic scattering at low temperatures}
\label{appx0}

Fig.~\ref{fig:invRT} shows a dip in $\tau^{-1}_{{\rm LO}}$ around the Fermi level $E_F$ at $T$=100~K.
This is a direct consequence of the Pauli exclusion principle for this inelastic process.
At low temperatures, most electrons occupy the states below $E_F$, which are nearly completely occupied. 
Due to the Pauli exclusion principle, they  preferentially scatter to the states above $E_F$, thus absorbing phonons. 
Since there are few thermally excited phonons at low temperatures, this scattering rate is very low. 
As $T$ increases, this dip disappears as electronic and phonon occupations broaden, as shown in Fig.~\ref{fig:invRT}(b) for $T$=600~K.

\section{Transport Properties at Room Temperature}
\label{appx1}

\begin{figure*}[h!]
  \centering
  \includegraphics{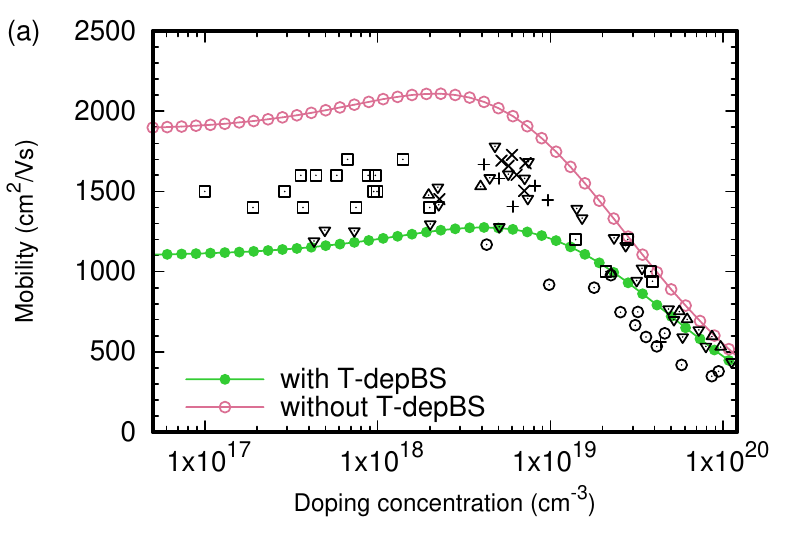}    
  \includegraphics{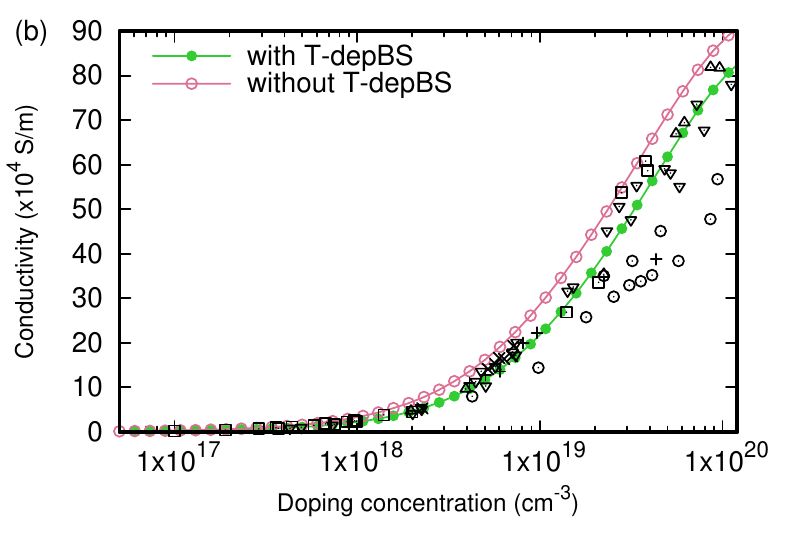} \\    
  \includegraphics{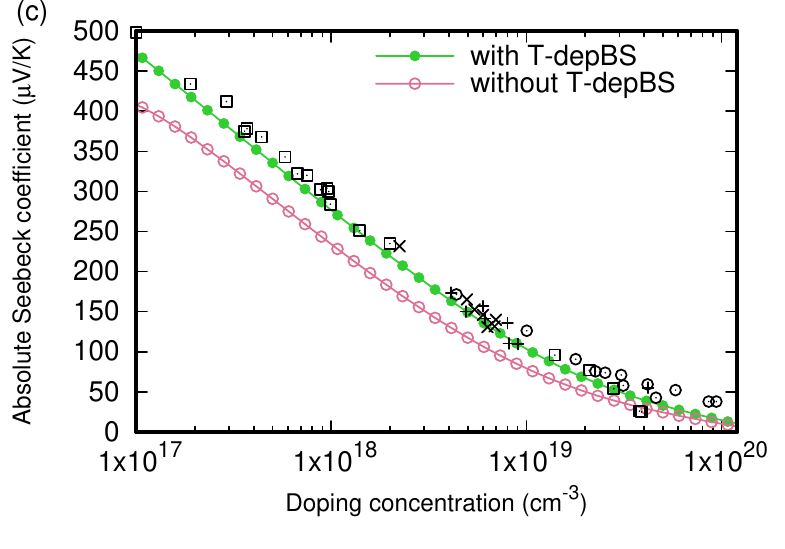}   
  \includegraphics{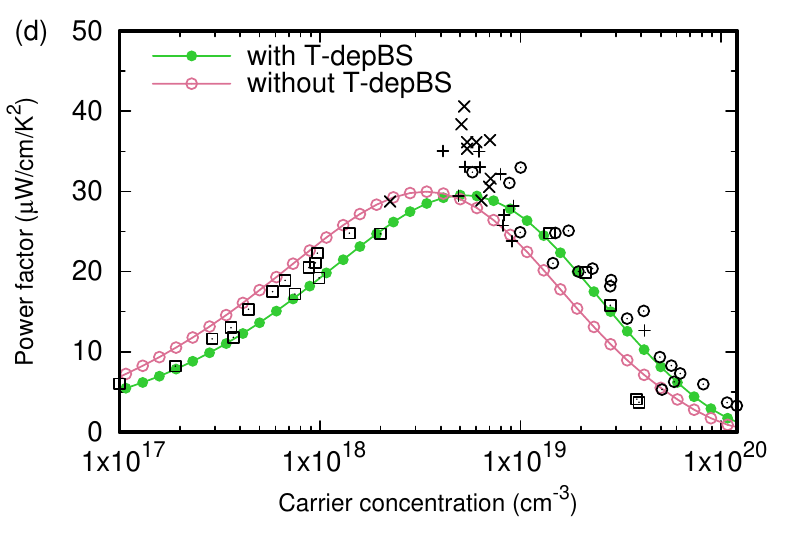}     
  \caption{Calculated and experimental dependence of 
    (a) the mobility, (b) the conductivity, (c) the absolute Seebeck coefficient 
    and (d) the power factor on the doping concentration at $T$=300~K. 
    Experimental data are shown in squares (Ref.~\cite{Harman1996}), 
    circles (Ref.~\cite{Snyder2014}), triangles up and down 
    (Ref.~\cite{Ravich1971b}), and crosses (Ref.~\cite{Vineis2008}). }    
  \label{fig:vsDoping}
\end{figure*}

The calculated thermoelectric transport 
properties as a function of the doping concentration for n-type PbTe at $300$~K including and 
excluding the temperature dependence of the electronic band structure (T-depBS) are 
illustrated in Fig.~\ref{fig:vsDoping}. The colored lines represent our calculations
while the symbols show the measurements from various experiments 
\cite{Ravich1971b,Harman1996,Snyder2014,Vineis2008}.
The calculated transport properties that account for the T-depBS are all in very good agreement
with the measurements. Including the T-depBS yields lower mobility and conductivity, and higher absolute 
Seebeck coefficient.

\bibliographystyle{elsarticle-num}
\bibliography{biblio}

\end{document}